\documentclass{article}
\usepackage{spconf}
\usepackage{amsmath}
\usepackage{amsthm}
\usepackage{amssymb}
\usepackage{amsfonts}
\usepackage{mathrsfs}
\usepackage{algorithmic,algorithm}
\usepackage{graphicx}
\usepackage{subfigure}
\usepackage{cite}
\usepackage{bm,epsfig,amsthm,url}
\usepackage{indentfirst}
\usepackage{epstopdf}
\usepackage{balance}
\usepackage{hyperref}
\usepackage{cleveref}

\title{Optimal Structure of Receive Beamforming for Over-the-Air Computation}
\name{Hongbin Zhu$^\dagger$, and Hua Qian$^\ast$
\thanks{This work was supported in part by Shanghai Sailing Program under Grant 23YF1402600.}}
\address{
$^\dagger$The FinTech Institute, Fudan University, Shanghai, 200433, China \\
$^\ast$Shanghai Advanced Research Institute, Chinese Academy of Sciences, Shanghai, 201210, China \\
Email: zhuhb@fudan.edu.cn, qianh@sari.cn\\
}

\begin{document}
\ninept

\maketitle
\begin{abstract}
We investigate fast data aggregation via over-the-air computation (AirComp) over wireless networks. In this scenario, an access point (AP) with multiple antennas aims to recover the arithmetic mean of sensory data from multiple wireless devices. 
To minimize estimation distortion, we formulate a mean-squared-error (MSE) minimization problem that considers joint optimization of transmit scalars at wireless devices, denoising factor, and receive beamforming vector at the AP. 
We derive closed-form expressions for the transmit scalars and denoising factor, resulting in a non-convex quadratic constrained quadratic programming (QCQP) problem concerning the receive beamforming vector. 
To tackle the computational complexity of the beamforming design, particularly relevant in massive multiple-input multiple-output (MIMO) AirComp systems, we explore the optimal structure of receive beamforming using successive convex approximation (SCA) and Lagrange duality. 
By leveraging the proposed optimal beamforming structure, we develop two efficient algorithms based on SCA and semi-definite relaxation (SDR). These algorithms enable fast wireless aggregation with low computational complexity and yield almost identical mean square error (MSE) performance compared to baseline algorithms. 
Simulation results validate the effectiveness of our proposed methods.
\end{abstract}

\section{Introduction}
Given the scarcity of spectrum resources and the demand for ultra-low latency, the conventional transmit-then-compute wireless transmission scheme falls short in meeting the requirements for fast wireless data aggregation. 
Fortunately, AirComp addresses this issue by enabling wireless devices to transmit information simultaneously over the same wireless channel\cite{yangkai,fang2020stochastic,DBLP:journals/tcom/FangJSZCL21}, resulting in fast wireless data aggregation within a single transmission interval. 
To achieve this, AirComp leverages the superposition property of multiple access channels (MACs).

Numerous studies have extensively investigated AirComp due to its potential for significant wireless data aggregation\cite{optimal_liu, optimal_huang, chen2018uniform, multi_function, Muti_modal, DBLP:conf/spawc/FangZZSZ21}.
The fundamental concept behind AirComp was initially proposed in a seminal study that explored functional computation in wireless sensor networks.
 In recent works, the authors of \cite{optimal_liu, optimal_huang} examined AirComp with multiple-input single-output (MISO) and developed a receive beamforming vector using a semi-definite relaxation (SDR) based successive convex approximation (SCA) algorithm.
 The authors of \cite{multi_function, Muti_modal}, on the other hand, developed transceiver designs for multi-function computation and multi-model sensing, respectively, for a multiple-input multiple-output (MIMO) AirComp system. 
Moreover, the authors of \cite{DBLP:conf/spawc/FangZZSZ21} devised an optimal receive beamforming algorithm for AirComp using the branch and bound (BnB) method.
Nevertheless, the high computational complexity associated with BnB makes it impractical for AirComp, particularly when the access point (AP) is equipped with multiple antennas.

This paper addresses the challenge of designing receive beamforming for an AirComp system equipped with a multi-antenna AP. SDR is a commonly employed numerical algorithm for approximating or approximating a globally optimal solution to this problem by relaxing it as a semi-definite program (SDP). 
However, SDR-based algorithms experience an increase in computational complexity as the problem size grows, leading to significant performance degradation. Consequently, SDR-based algorithms are ill-suited for multi-antenna APs, especially in the case of MIMO systems with large-scale antenna arrays.

To overcome these limitations, the authors of \cite{chen2018uniform} introduced the use of successive convex approximation (SCA) for the receive beamforming design in AirComp. SCA employs a series of convex approximations to iteratively address the original problem. Compared to SDR, SCA offers improved performance with reduced complexity. Nevertheless, SCA still entails a high computational load when applied to AirComp systems featuring a large number of antennas.

Motivated by the necessity for low-complexity algorithms in the design of receive beamforming for AirComp, our objective is to develop efficient solutions. 
Inspired by \cite{DBLP:journals/tsp/DongW20}, by leveraging both SCA and Lagrangian duality, we derive the optimal structure for receive beamforming in AirComp, reducing the dimension of the problem from the number of antennas to the number of wireless devices. Leveraging this proposed optimal structure, we devise two efficient algorithms based on SDR and SCA, respectively. These algorithms significantly reduce the computational complexity associated with receive beamforming while maintaining the same level of mean square error (MSE) performance as achieved by the original SDR and SCA.

\emph{Notations}: 
Boldface upper-case, boldface lower-case, and lower-case letters denote matrices, vectors, and scalars, respectively.
The imaginary unit of a complex number is denoted as $\mathbf{j}$.
The conjugate transpose of a matrix or a vector is represented as $(\bm \cdot)^{\sf{H}}$.
The $l_2$ norm operator is denoted by $\|\cdot\|$.
The real part, imaginary part, absolute value, and argument of a scalar are denoted by $\operatorname{Re}\{\bm\cdot\}$, $\operatorname{Im}\{\bm \cdot\}$, $|\cdot|$, and $\operatorname{arg}(\cdot)$, respectively.
The expectation of a random variable is denoted as $\mathbb{E}\left[\bm \cdot\right]$.

\section{System Model and Problem Formulation}

We investigate AirComp in a wireless system comprising $K$ single-antenna wireless devices and an AP with $N$ antennas.
The objective of the AP is to retrieve the arithmetic mean of the sensory data from all wireless devices. 
Denote the index set of devices as $\mathcal{K}=\{1,2,\ldots, K\}$.
The transmit signal of device $k$ is denoted as $s_k = \varphi_k(z_k)$, where $\varphi_k(\cdot)$ represents the specific pre-processing function and $z_k\in\mathbb{C}$ represents the representative information-bearing data at device $k$.
We assume that $\{s_k\}_{k=0}^{K}$ are independent, have zero mean, and unit power, i.e., $\mathbb{E}[s_{k}s_{k}^{\sf H}] = 1$, and $\mathbb{E}[s_{k}s_{j}^{\sf H}] = 0, \forall  k \neq j$\cite{optimal_huang}.
The AP estimates the target function $g=\sum_{k\in\mathcal{K}}s_k$ to obtain the arithmetic mean of the sensory data from all the connected wireless devices, i.e., $\frac{1}{K}\sum_{k\in\mathcal{K}}z_k$.
By synchronizing the transmission timing of each device, the signals transmitted by all wireless devices can be aligned. 
Consequently, we can represent the received signal at the AP as follows
\begin{align}
\bm{y}=\sum_{k\in\mathcal{K}}\bm{h}_{k}{w}_ks_k+\bm{n},
\end{align}
where $w_k\in\mathbb{C}$ represents the transmit scalar of device $k$, $ \mathbf h_{k}\in\mathbb{C}^{N\times1}$ represents the channel coefficient vector of the link from device $k$ to AP,
 and $ \bm n\sim\mathcal{CN}(0, \sigma^2\bm I_N) $ represents the additive white Gaussian noise (AWGN) with zero mean and variance $\sigma^2$. 
In practice, the maximum transmit power is limited, i.e., $|w_k|^2 \leq P, \forall k$.
After the received signal combination at AP, the estimated function is as follows\cite{chen2018uniform, DBLP:conf/spawc/FangZZSZ21}
\begin{align}
\hat{g}&={1\over{\sqrt \eta}}{\bm{m}^{\sf{H}}\bm{y}} =\!{1\over{\sqrt \eta}}{\bm{m}}^{\sf{H}}\sum_{k\in\mathcal{K}}\bm{h}_{k} {w}_ks_k +{1\over{\sqrt \eta}}\bm{m}^{\sf{H}}\bm{n},
\end{align} 
where $\bm{m}\in\mathbb{C}^N$ and $\eta$ denote the receive beamforming vector and the denoising factor at AP, respectively. 
  
We use MSE criterion as the metric to assess the distortion between $\hat{g}$ and $g$, which is
\begin{align}
{\sf{MSE}}(\hat{g}, g)=\mathbb{E}\left(|\hat{g}-g|^2\right) = \!\! \sum_{k\in\mathcal{K}}\left|   \frac{{{\bm{m}}^{\sf{H}}\bm{h}_{k}{w}_k}}{\sqrt{\eta}} -1\right|^2 \!+\! \frac{\sigma^2\|\bm{m}\|^2}{\eta}\nonumber . 
\end{align}
After determining receive beamforming vector $\bm{m}$, the optimal transmit scalars that realize the minimization of MSE are as follows \cite{yangkai,chen2018uniform}
\begin{align}\label{a}
w_k^{\star}=\sqrt{\eta}{{(\bm{m}^{\sf{H}}\bm{h}_{k})^{\sf{H}}}\over{\|\bm{m}^{\sf{H}}\bm{h}_{k}\|^2}},\forall k. 
\end{align}
Due to the transmit power constraint, we can express $\eta$ as \begin{align}\label{b}
\eta=P\min_{k\in\mathcal{K}} \|\bm{m}^{\sf{H}}\bm{h}_{k}\|^2.
\end{align}
With \eqref{a} and \eqref{b}, we can further rewrite MSE as
\begin{align}
{\sf{MSE}}={{\|\bm{m}\|^2\sigma^2}\over{\eta}}
={{\|\bm{m}\|^2\sigma^2}\over{P\min_{k\in\mathcal{K}} \|\bm{m}^{\sf{H}}\bm{h}_{k}\|^2}}.
\end{align}
Hence, we endeavor to find receive beamforming  vector $ \bm m $ as follows:  
\begin{equation}\label{eq:ori}
        \begin{split}
                \underset{\bm m}{\min} \left({{\|\bm{m}\|^2\sigma^2}\over{P\min_{k\in\mathcal{K}} \|\bm{m}^{\sf{H}}\bm{h}_{k}\|^2}}\right).
        \end{split}
\end{equation}
From \cite{chen2018uniform}, we can equivalently transform problem \eqref{eq:ori} to the following non-convex quadratic constrained quadratic programming (QCQP) problem
\begin{equation*}
\begin{aligned}
        \mathcal{P}_o :\quad
                \underset{\bm m}{\min}  &\quad \|\bm m\|^2\quad\quad\quad\quad\quad\quad\quad\quad \\
        \end{aligned}
\end{equation*}
\begin{equation}\label{equ:original_constraint} 
 \quad \text{s.t.} \quad\|\bm m^{\sf H}\bm{h}_{k}\|^2 \geq 1, ~  \forall k.	
\end{equation}


\section{Optimal receive beamforming strucure}\label{sec:optimal}
In this section, we investigate the optimal receive beamforming structure using the SCA algorithm and Lagrangian duality.
With SCA, by introducing the auxiliary vector $\bm{z}\in \mathbb{C}^{N\times1}$, we obtain the following optimization problem, i.e.,
\begin{equation*}\label{pro:original_sca} 
\begin{aligned}
        \mathcal{P}_{\text{SCA}}(\bm{z}) :
                \underset{\bm m}{\min}  &\quad \|\bm m\|^2\\
                \text{s.t.} &\quad |\bm z^{\sf H}\bm{h}_{k}|^2-2\operatorname{Re}\{\bm{m}^{\sf H}\bm{h}_k \bm{h}_k^{\sf H}\bm{z}\} \leq -1, ~  \forall k.
        \end{aligned}
\end{equation*}
$\mathcal{P}_{\text{SCA}}$ is a convex approximation of $\mathcal{P}_o$.

By replacing non-convex constraint \eqref{equ:original_constraint} with convex constraint, $\mathcal P_{\text{SCA}}(\bm z)$ becomes convex.
Therefore, we employ standard SCA algorithm to tackle $\mathcal P_{\text{SCA}}(\bm z)$.
In particular, SCA can guarantee the convergence of $\mathcal P_o$ to a stationary point\cite{marks1978general}.
When choosing the initial point $\bm m^{(0)}$ appropriately, i.e., $\bm z^{(0)}$ is at the vicinity of the global optimal solution $\bm m^o$, the solution of $\mathcal{P}_{\text{SCA}}(\bm z)$ obtained by SCA is guaranteed to  converge to global optimal solution of $\mathcal P_o$. 
In this paper, we leverage SDR to obtain $\bm z^{(0)}$ for $\mathcal P_{\text{SCA}}(\bm z)$.

Next, we present the steps to obtain the optimal receive beamforming of AirComp.
As $\mathcal{P}_{\text{SCA}}(\bm z)$ is convex, then Slater's condition holds.
Therefore, we can obtain its optimal solution from its Lagrange dual domain.
The Lagrangian for $\mathcal{P}_{\text{SCA}}(\bm z)$ is as follows
\begin{align}\label{equ:Lagrangian}
	\mathcal{L}(\bm z, \bm m, \bm \lambda)  = &\sum_{k=1}^K\lambda_k\left( |\bm z^{\sf H}\bm{h}_{k}|^2-2\operatorname{Re}\{\bm{m}^{\sf H}\bm{h}_k \bm{h}_k^{\sf H}\bm{z}\}+1\right) \notag \\ 
	& +\|\bm m\|^2,
\end{align}
where $\lambda_k$ represents the Lagrange multiplier associated with constraint \eqref{equ:original_constraint} for device $k$, and $\bm \lambda=\left[\lambda_1, \lambda_2, \ldots, \lambda_K\right]$.
The Lagrange dual problem for $\mathcal{P}_{\text{SCA}}(\bm z)$ is as follows 
\begin{equation}
	\mathcal{D}_{\text{SCA}}(\bm z): \quad
                \underset{\bm \lambda}{\max}\; g(\bm z, \bm \lambda)\quad \text{s.t.}\; \bm \lambda \succcurlyeq \bm 0,  
\end{equation}
 where 
 \begin{equation}
 	g(\bm z, \bm \lambda) = \underset{\bm m}{\min}\; \mathcal{L}(\bm z, \bm m, \bm \lambda).
 \end{equation}
Reordering \eqref{equ:Lagrangian}, we have 
\begin{align}\label{equ:Lagrangian_reorder}
	\mathcal{L}(\bm z, \bm m, \bm \lambda)  = & \|\bm m\|^2-\sum_{k=1}^K2\operatorname{Re}\{\lambda_k\bm{m}^{\sf H}\bm{h}_k \bm{h}_k^{\sf H}\bm{z}\}\notag \\ 
	& +\sum_{k=1}^K\lambda_k\left( |\bm z^{\sf H}\bm{h}_{k}|^2+1\right).
\end{align}
Accordingly, we can equivalently transform the optimization problem \eqref{equ:Lagrangian} to 
\begin{equation}\label{equ:L_min}
	 \underset{\bm m}{\min}\;  \|\bm m\|^2-\sum_{k=1}^K2\operatorname{Re}\{\lambda_k\bm{m}^{\sf H}\bm{h}_k \bm{h}_k^{\sf H}\bm{z}\}.
\end{equation}
Since \eqref{equ:L_min} is convex, we can leverage KKT conditions to obtain its optimal solution in closed form.
\newtheorem*{Pro:1}{Proposition 1}
\begin{Pro:1}
The optimal solution of $\mathcal{P}_{\text{SCA}}$ is as follows
\begin{equation}
	 \bm m^{\ast}(\bm z)= \sum_{k=1}^K\lambda_k^{\ast}\bm{h}_k^{\sf H} \bm z\bm{h}_k, 
\end{equation}
where $\bm \lambda^\ast = [\lambda_1^\ast, \lambda_2^\ast, \ldots, \lambda_K^\ast]$ represents the optimal dual solution for $\mathcal{D}_{\text{SCA}}(\bm z)$.
\begin{proof}[Proof]
Given $\bm z$, we use $J(\bm z, \bm m)$ to denote the objective function in \eqref{equ:L_min}.
By KKT condition, at the optimality of $\mathcal{P}_{\text{SCA}}(\bm z)$, the gradient of $J(\bm z, \bm m)$ w.r.t. $\bm m$ is
 \begin{equation}
 \nabla_{\bm m}J(\bm z, \bm m) = \bm{m}^{\ast}- \sum_{k=1}^K\lambda_k^{\ast}\bm{h}_k \bm{h}_k^{\sf H} \bm z = \bm 0, 
 \end{equation}
 and we obtain $\bm m^{\ast}(\bm z)= \sum_{k=1}^K\lambda_k^{\ast}\bm{h}_k^{\sf H} \bm z\bm{h}_k$.
\end{proof}
\end{Pro:1}

Examining the optimal solution $\bm m^{\ast}(\bm z)$ in $\textbf{Proposition 1}$, we note that $\bm m^{\ast}$ depends on $\bm z$.
This observation denotes that the optimal solution $\bm m^{\ast}(\bm z)$ for $\mathcal P_{\text{SCA}}(\bm z)$ is updated accordingly when SCA algorithm iteratively updates $\bm z$, while the structure of $\bm m^{\ast}(\bm z)$ remains the same.
Hence, if $\bm z \rightarrow \bm m^o$, we obtain the optimal solution for $\mathcal P_o$.
\newtheorem*{Theo:1}{Theorem 1}
\begin{Theo:1}
The optimal receive beamforming solution for AirComp beamforming problem $\mathcal{P}_o$ is given by
\begin{equation}\label{equ:optimal_s}
	\bm m^o = \mathbf{H} \bm a^o,
\end{equation}
where $\mathbf H = [\bm h_1,\bm{h}_2, \ldots, \bm h_K]$, $a_k^o = \lambda_k^o \mathbf H_k^H \bm m^o$, $\bm a^o = [a^o_1, a^o_2,\ldots,$\\
$a^o_K]$.
\begin{proof}
The SCA iteration is guaranteed to converge to a stationary point.
Therefore, if $\bm z^{(0)}$ is initialized at the vicinity of the global optimal solution, SCA algorithm will converge to the global optimal solution, i.e., $\bm z \rightarrow \bm m^o$.
Also, as $\bm z \rightarrow \bm m^o$, the optimal $\bm \lambda^\ast$ for $\mathcal{D}(\bm z)$ converges to $\bm \lambda^o$ for $\mathcal{D}(\bm m^o)$.
Therefore, we have $\bm m^o = \sum_{k=1}^K\lambda_k^{o}\bm{h}_k^{\sf H} \bm m^o\bm{h}_k = \mathbf{H} \bm a^o$.
\end{proof}
\end{Theo:1}

Note that the optimal solution $\bm m^o$ in \eqref{equ:optimal_s} is expressed in semi-closed form, where $\bm a^o$ should be determined numerically.
Computing the optimal $\bm m^o$ is still challenging because $\mathcal P_o$ is NP-hard.

\section{Proposed efficient algorithms}\label{sec:algorithms}
In this section, we first provide a detailed description of efficient algorithms to compute $\bm a^o$.
We then discuss the computation complexity of the developed algorithms.

We first define $\bm f_k =  \mathbf H^{\sf H} \bm{h}_{k}$.
By leveraging the optimal structure of $\bm m^o$ in \eqref{equ:optimal_s}, we can recast the receive beamforming problem $\mathcal{P}_o$ into a weight optimization as follows
\begin{equation*}
\begin{aligned}
        \mathcal{P}_1 :\quad
                \underset{\bm a}{\min}  &\quad \|\mathbf H \bm a\|^2\quad\quad\quad\quad\quad\quad\quad\quad\quad \\
        \end{aligned}
\end{equation*}
\begin{equation}\label{equ:P1_constraint} 
  \text{s.t.} \quad |\bm a^{\sf H}  \bm{f}_{k}|^2 \geq 1, ~  \forall k.	\quad\;
\end{equation}

Similar to the constraint in $\mathcal P_o$, the constraint \eqref{equ:P1_constraint} is non-convex. 
Hence, $\mathcal P_1$ is still NP-hard.
However, the key difference here is that the beamforming vector $\bm m$ in $\mathcal{P}_o$ is of size $N$. 
In contrast, the weight vector $\bm a$ for the weight optimization problem $\mathcal{P}_1$ is of size $K$, which no longer depends on $N$.
This is particularly appealing to massive MIMO systems with $K \ll N$.
Solving $\mathcal P_1$ with much reduced size instead of $\mathcal P_o$ reduces the computation cost significantly.

Next, we propose to employ two approaches, i.e., SDR and SCA, to compute the weight vector $\bm a$ for $\mathcal P_1$.

$1)\; \emph{The SDR algorithm:}$ Defining $\mathbf X=\bm a \bm a^{\sf H}$, $\bm H_k = \bm h_k \bm h_k^{\sf H}$, $\mathbf D = \mathbf H^{\sf H} \mathbf H$, and dropping the rank-one constraint on $\mathbf X$, $\mathcal P_1$ is relaxed to the following SDP problem
 \begin{equation*}
\begin{aligned}
        \mathcal{P}_{1\text{SDR}} :\quad
                \underset{\mathbf X}{\min}  &\quad \text{tr} (\mathbf D \mathbf X )\quad\quad\quad\quad\quad\quad\quad\quad\quad\quad \\
        \end{aligned}
\end{equation*}
\begin{equation*}
  \text{s.t.} \quad \text{tr}( \bm{f}_{k}\bm{f}_{k}^{\mathsf H}\mathbf X) \geq 1, ~  \forall k, 
\end{equation*}
\begin{equation*}
	\mathbf X\succeq 0. \quad\quad\quad\;
\end{equation*}

$2)\; \emph{The SCA algorithm:}$ Define $\bm f_k = \mathbf H^{\sf H}\bm h_k$, and leverage the auxiliary variable $\bm y \in \mathbb{C}^{K\times1}$.
 By applying convex approximation to constraint \eqref{equ:P1_constraint} in $\mathcal P_1$, we have the following convex optimization problem for any given $\bm y$
 \begin{equation*}
\begin{aligned}
        \mathcal{P}_{1\text{SCA}}(\bm y) :\quad
                \underset{\bm a}{\min}  &\quad  \|\mathbf H \bm a \|^2\quad\quad\quad\quad\quad\quad\quad\quad\quad\quad\quad  \\
        \end{aligned}
\end{equation*}
\begin{equation}\label{equ:P1SCA_constraint} 
  \quad\;\quad\quad \text{s.t.} \quad 2\operatorname{Re} \{\bm a^{\sf H} \bm f_k \bm f_k^{\sf H} \bm y\}-|\bm y^{\sf H}\bm f_k|^2 \geq 1, ~  \forall k. 
\end{equation} 

The SCA algorithm requires that the initial $\bm y^{(0)}$ is feasible for $\mathcal P_{1\text{SCA}}(\bm y)$.
To ensure this, we employ the solution $\bm a^{\text{SDR}}$ of $\mathcal P_{1\text{SDR}}$ for the initialization of $\mathcal P_{1\text{SCA}}(\bm y)$, i.e., $\bm y^0 = \bm a^{\text{SDR}}$.
The solution $\bm a^{\text{SDR}}$ provides a good initial point close to the optimum of $\mathcal P_o$, which in turn will fasten the convergence of SCA algorithm.
Compared to original problem $\mathcal P_o$, the problem size of $\mathcal P_{1\text{SCA}}$ is smaller, computing $\bm a^{\text{SDR}}$ is fast even for large $N$ and we will verify it in Section \ref{sec:simulation}.
Compared to directly solving $\mathcal P_o$ by SDR, adopting SDR to solve $\mathcal P_1$ can significantly reduce the computation complexity.
Specifically, by employing the typical interior point methods, the worst case complexity of solving $\mathcal P_1$ is $\mathcal{O}((K^2)^{3.5})$ \cite{DBLP:journals/tsp/SidiropoulosDL06}, while the worst case complexity of directly solving $\mathcal P_o$ via SDR is $\mathcal{O}((N^2)^{3.5})$.
On the other hand, the computation complexity in each iteration is $\mathcal{O}(K^{3})$ by leveraging SCA with typical interior-point method to tackle $\mathcal P_{1\text{SCA}}(\bm y)$.
The computation complexity in each iteration is $\mathcal{O}(N^{3})$ to directly solve $\mathcal P_o$ via SCA.
This complexity analysis shows that we can significantly reduce the computation complexity with the help of the proposed optimal structure.

\section{Simulation Results}\label{sec:simulation}
In this section, we present simulation results of the proposed algorithm for AirComp in wireless networks.
We consider a three-dimensional setting where the AP is located at coordinates $(0,0,20)$, and devices are uniformly distributed within a circular region centered at $(120,~ 20,~ 0)$ meters with a radius of $20$ meters.
The antennas at the AP are arranged as a uniform linear array. 
We consider both large-scale fading and small-scale fading for the wireless channel.
We model the distance-dependent large-scale fading as $T_0(d/d_0)^{-\alpha}$, 
where $T_0$ is the path loss at the reference distance $d_0 = 1$ meter, $d$ denotes the distance between transmitter and receiver, and $\alpha$ is the path loss exponent.
Additionally, we model small-scale fading as Rician fading with Rician factor $\beta$.
All simulation results are obtained by averaging over $128$ channel realizations.
Unless specified otherwise, we set $\alpha = 3$, $T_0 = -30 $ dB, $\beta = 3$, $P = 30$ dBm, $\sigma^2 = -100$ dBm, and $\epsilon = 10^{-5}$.

 We evaluate the performance of the proposed algorithms utilizing the optimal beamforming structure $\bm m^o$ in \eqref{equ:optimal_s} for $\mathcal P_o$. 
 We employ both SDR and SCA algorithms in Section \ref{sec:algorithms} to compute the weight vector $\bm a$ for $\mathcal P_{1\text{SDR}}$ and $\mathcal P_{1\text{SCA}}$, which we refer to as SDR-Opt and SCA-Opt, respectively, for clarity. 
 To demonstrate the effectiveness of our approach, we compare our algorithms to the following baseline methods: 1) Direct SDR: $\mathcal P_o$ is solved directly via SDR with Gaussian randomization; 2) Direct SCA: $\mathcal P_o$ is solved directly via SCA, taking the solution from direct SDR as the initial point.

\begin{figure}[!t]
	\centering
	\includegraphics[width=70mm, height = 5.4cm]{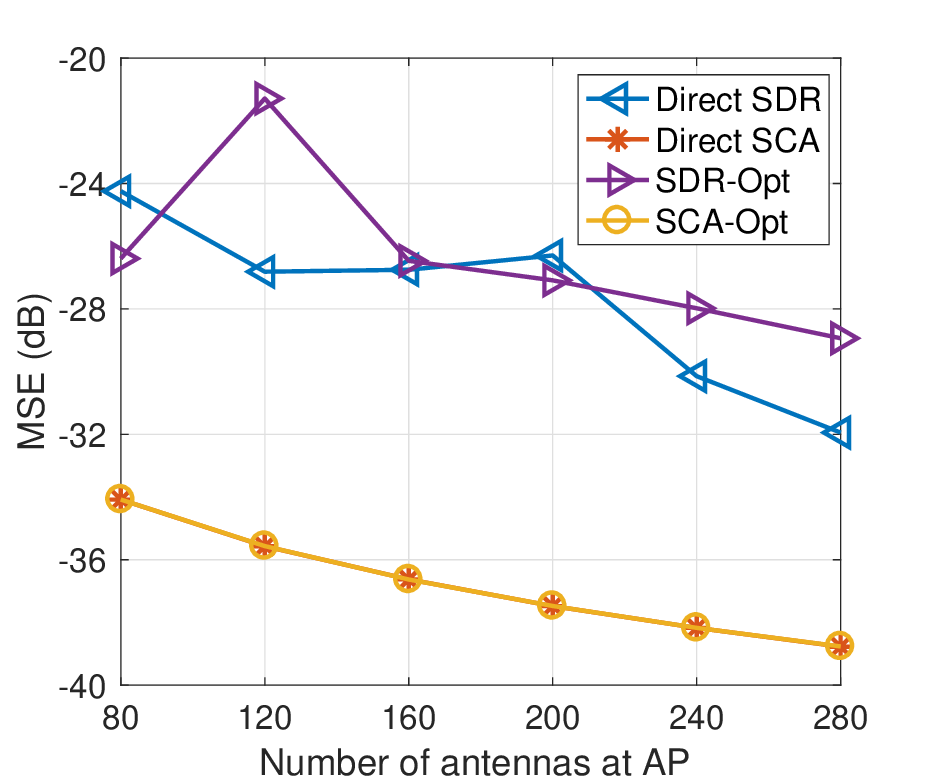}
	\caption{MSE versus $N$ at AP when $K = 10$.}
	\label{fig:antennas_no}
	\vspace{-3mm}
\end{figure}
\begin{figure}[!t]
	\centering
	\includegraphics[width=70mm, height = 5.4cm]{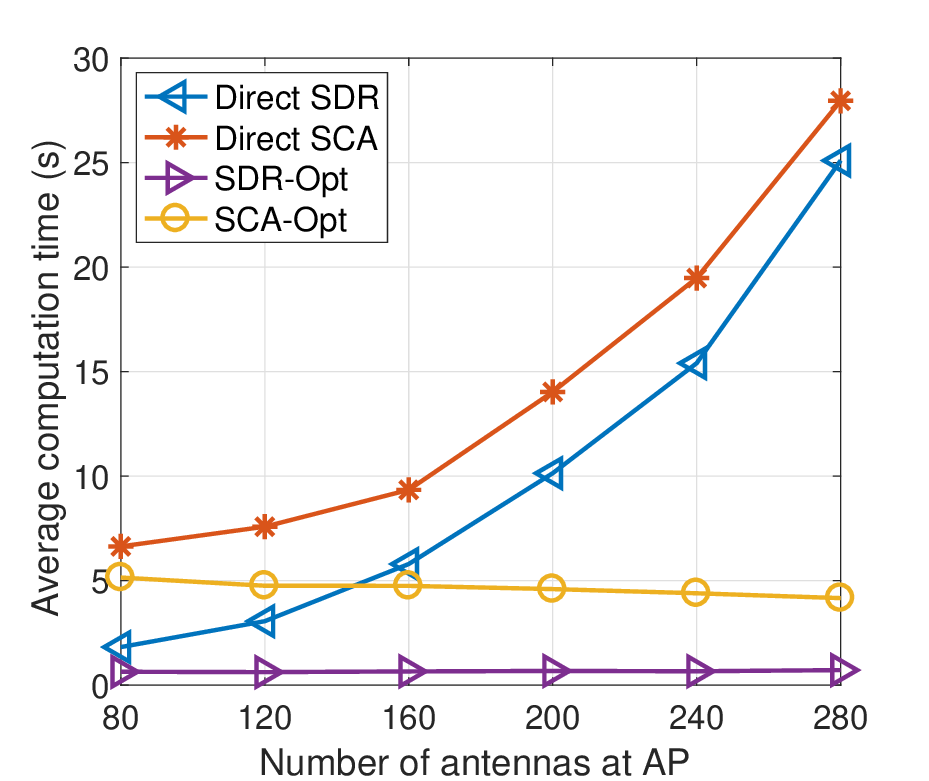}
	\caption{Average computation time versus $N$ when $K = 10$.}
	\label{fig:antennas_time}
	\vspace{-3mm}
\end{figure}



Fig. \ref{fig:antennas_no} depicts the effect of the number of antennas at the AP on the MSE with a fixed number of devices $K = 10$. 
From the SCA-based algorithms, we can observe that as the number of antennas increases, the MSE of AirComp monotonically decreases due to the increased diversity gain. 
Compared to the direct SDR and SDR-Opt, SCA-Opt exhibits a better MSE performance that is almost identical to the direct SCA. 
However, there is a significant performance gap between SCA-Opt and SDR-based algorithms, as the latter is not optimized for the AirComp system.

Fig. \ref{fig:antennas_time} depicts the impact of the number of antennas at AP on the average computing time with a fixed $K = 10$. 
The results show that the average computation time of SDR-Opt (SCA-Opt) is lower than that of direct SDR (SCA). 
Moreover, the number of antennas at AP has negligible impact on the average computation time of SDR-Opt and SCA-Opt since their computation complexity is only dependent on $K$. 
On the other hand, SCA based algorithms incur more computing time than SDR based algorithms since they require an initial solution point from SDR based algorithms. 
The experimental results in Fig. \ref{fig:antennas_time} demonstrate that the proposed optimal structure can significantly reduce computation time without degrading the MSE performance. 

\begin{figure}[!t]
	\centering
	\includegraphics[width=70mm, height =5.4cm]{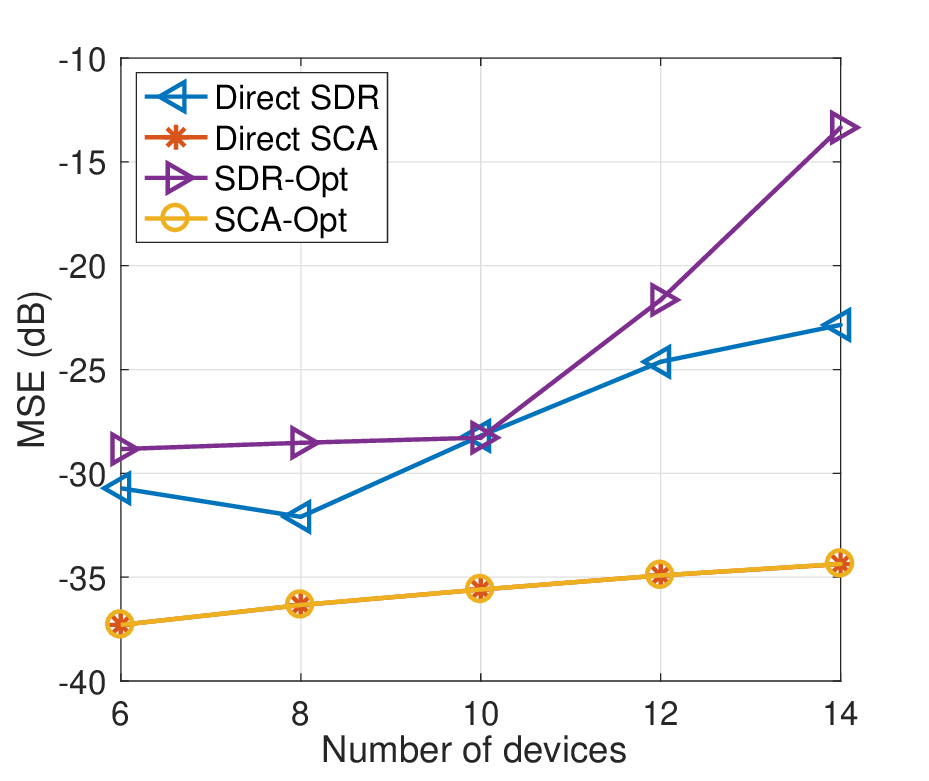}
	\caption{MSE versus $K$ when $N = 120$.}
	\label{fig:devices_no}
	\vspace{-3mm}
\end{figure}
\begin{figure}[!t]
	\centering
	\includegraphics[width=70mm, height =5.4cm]{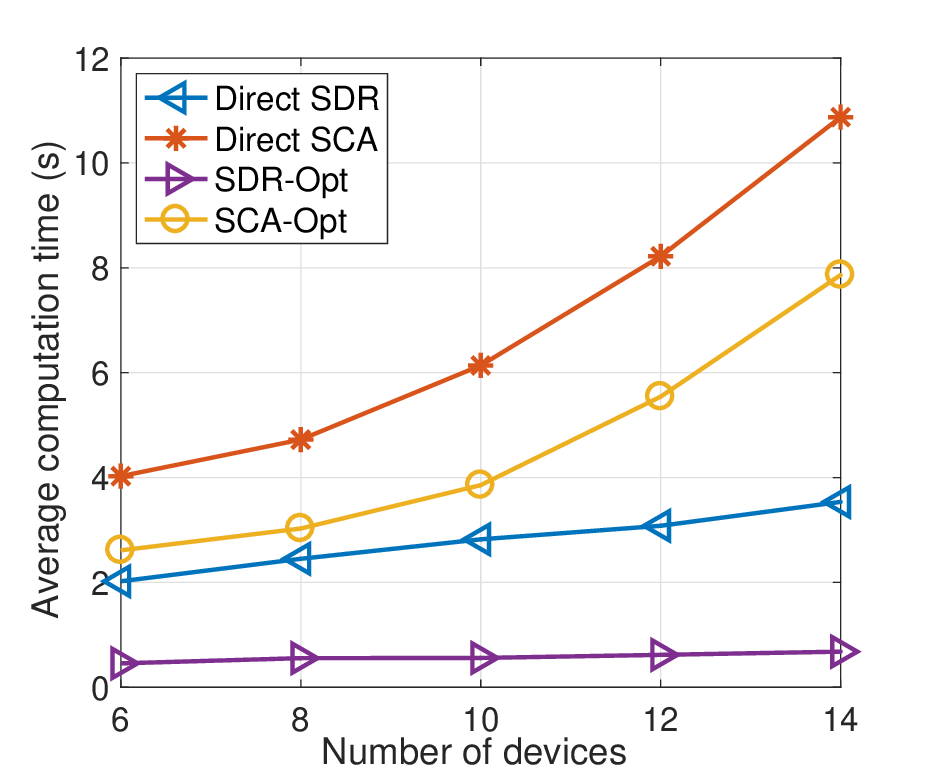}
	\caption{Average computation time versus $K$ when $N = 120$.}
	\label{fig:devices_time}
	\vspace{-3mm}
\end{figure}

Fig. \ref{fig:devices_no} displays the relationship between MSE and the number of wireless devices, where the number of antennas at the AP is set to $10$. 
We observe that SCA-Opt achieves better MSE performance than direct SDR-based algorithms, which is almost identical to direct SCA. 
Moreover, as the number of wireless devices increases, the quality of the solutions for SCA-based algorithms deteriorates noticeably. 
This is because the MSE performance is determined by the worst wireless link between the device and the AP, and it decreases with the increase in the number of wireless devices.

Fig. \ref{fig:devices_time} illustrates the impact of the number of devices on the average computing time, with a fixed number of antennas at the AP ($N = 120$). 
Comparing SDR-Opt (SCA-Opt) with direct SDR (direct SCA), we observe that SDR-Opt (SCA-Opt) incurs less computation time. 
Additionally, the average computation time of SCA-Opt increases with the increase of the number of devices, owing to the increment in computation complexity with $K$.
The results in Fig. \ref{fig:devices_time} demonstrate that the proposed optimal structure could significantly reduce the computation time for the receiver beamforming design of AirComp.

\section{Conclusions}
In this paper, we investigated the joint design of transmit scalars, denoising factor, and receive beamforming vector for the AirComp system. 
We derived closed-form expressions for the transmit scalars and denoising factor, resulting in a non-convex QCQP problem with respect to the receive beamforming vector at the AP. 
Through the utilization of the SCA numerical algorithm and Lagrange duality, we obtained the optimal receive beamforming structure for the AirComp system. 
Notably, the optimal beamforming structure remains independent of the system's large parameter N, which is particularly advantageous for massive MIMO AirComp systems.
Leveraging the optimal receive beamforming structure, we developed two highly efficient algorithms, named SDR-Opt and SCA-Opt. 
Compared to baseline algorithms, these newly developed efficient algorithms require much lower computation complexity while achieving almost identical MSE performance.

\balance
\bibliographystyle{IEEEtran}  
\bibliography{reference.bib}

\begin{thebibliography}{10}
\providecommand{\url}[1]{#1}
\csname url@samestyle\endcsname
\providecommand{\newblock}{\relax}
\providecommand{\bibinfo}[2]{#2}
\providecommand{\BIBentrySTDinterwordspacing}{\spaceskip=0pt\relax}
\providecommand{\BIBentryALTinterwordstretchfactor}{4}
\providecommand{\BIBentryALTinterwordspacing}{\spaceskip=\fontdimen2\font plus
\BIBentryALTinterwordstretchfactor\fontdimen3\font minus \fontdimen4\font\relax}
\providecommand{\BIBforeignlanguage}[2]{{%
\expandafter\ifx\csname l@#1\endcsname\relax
\typeout{** WARNING: IEEEtran.bst: No hyphenation pattern has been}%
\typeout{** loaded for the language `#1'. Using the pattern for}%
\typeout{** the default language instead.}%
\else
\language=\csname l@#1\endcsname
\fi
#2}}
\providecommand{\BIBdecl}{\relax}
\BIBdecl

\bibitem{yangkai}
K.~{Yang}, T.~{Jiang}, Y.~{Shi}, and Z.~{Ding}, ``Federated learning via over-the-air computation,'' \emph{IEEE Trans. Wireless Commun.}, vol.~19, no.~3, pp. 2022--2035, Mar. 2020.

\bibitem{fang2020stochastic}
W.~Fang, M.~Fu, K.~Wang, Y.~Shi, and Y.~Zhou, ``Stochastic beamforming for reconfigurable intelligent surface aided over-the-air computation,'' in \emph{Proc. IEEE Global Commun. Conf. (Globecom)}, Dec. 2020.

\bibitem{DBLP:journals/tcom/FangJSZCL21}
W.~Fang, Y.~Jiang, Y.~Shi, Y.~Zhou, W.~Chen, and K.~B. Letaief, ``Over-the-air computation via reconfigurable intelligent surface,'' \emph{{IEEE} Trans. Commun.}, vol.~69, no.~12, pp. 8612--8626, 2021.

\bibitem{optimal_liu}
W.~{Liu}, X.~{Zang}, Y.~{Li}, and B.~{Vucetic}, ``Over-the-air computation systems: Optimization, analysis and scaling laws,'' \emph{IEEE Trans. Wireless Commun.}, vol.~19, no.~8, pp. 5488--5502, Aug. 2020.

\bibitem{optimal_huang}
X.~{Cao}, G.~{Zhu}, J.~{Xu}, and K.~{Huang}, ``Optimized power control for over-the-air computation in fading channels,'' \emph{IEEE Trans. Wireless Commun.}, vol.~19, no.~11, pp. 7498--7513, Nov. 2020.

\bibitem{chen2018uniform}
L.~{Chen}, X.~{Qin}, and G.~{Wei}, ``A uniform-forcing transceiver design for over-the-air function computation,'' \emph{IEEE Wireless Commun. Lett.}, vol.~7, no.~6, pp. 942--945, Dec. 2018.

\bibitem{multi_function}
L.~{Chen}, N.~{Zhao}, Y.~{Chen}, F.~R. {Yu}, and G.~{Wei}, ``Over-the-air computation for {IoT} networks: Computing multiple functions with antenna arrays,'' \emph{IEEE Internet Things J.}, vol.~5, no.~6, pp. 5296--5306, Jun. 2018.

\bibitem{Muti_modal}
G.~{Zhu} and K.~{Huang}, ``{MIMO} over-the-air computation for high-mobility multimodal sensing,'' \emph{IEEE Internet Things J.}, vol.~6, no.~4, pp. 6089--6103, Aug. 2019.

\bibitem{DBLP:conf/spawc/FangZZSZ21}
W.~Fang, Y.~Zou, H.~Zhu, Y.~Shi, and Y.~Zhou, ``Optimal receive beamforming for over-the-air computation,'' in \emph{Proc. {IEEE} International Workshop on Signal Process. Advances in Wireless Commun. {(SPAWC)}}, pp. 61--65, Sep. 2021.

\bibitem{DBLP:journals/tsp/DongW20}
M.~Dong and Q.~Wang, ``Multi-group multicast beamforming: Optimal structure and efficient algorithms,'' \emph{{IEEE} Trans. Signal Process.}, vol.~68, pp. 3738--3753, 2020.

\bibitem{marks1978general}
B.~R. Marks and G.~P. Wright, ``A general inner approximation algorithm for nonconvex mathematical programs,'' \emph{Oper. Res.}, vol.~26, no.~4, pp. 681--683, 1978.

\bibitem{DBLP:journals/tsp/SidiropoulosDL06}
N.~D. Sidiropoulos, T.~N. Davidson, and Z.~Luo, ``Transmit beamforming for physical-layer multicasting,'' \emph{{IEEE} Trans. Signal Process.}, vol.~54, no. 6-1, pp. 2239--2251, 2006.

\end{thebibliography}

\end{document}